\documentclass[useAMS,usenatbib]{mn2e}
\usepackage{graphicx,amssym}
\newcommand{\bc}{\begin{center}}
\newcommand{\ec}{\end{center}}

\title[Galaxy Formation: WMAP1 vs WMAP3] {The Dependence of Galaxy
      Formation on Cosmological Parameters: Can we distinguish the
      WMAP1 and WMAP3 Parameter Sets?}  
      \author[J. ~Wang et al.]  {Jie Wang\thanks{Email:
      wangjie@mpa-garching.mpg.de}, ~Gabriella De Lucia, ~Manfred
      G. Kitzbichler,~Simon D.~M.~White \\
      Max--Planck--Institut f\"ur Astrophysik,
      Karl--Schwarzschild--Str. 1, D-85748 Garching, Germany}
\begin{document}

\date{Accepted 2007 ???? ??. 
      Received 2007 ???? ??; 
      in original form 2007 ???? ??}

\pagerange{\pageref{firstpage}--\pageref{lastpage}} 
\pubyear{2007}

\maketitle
\label{firstpage}
\begin{abstract}
 We combine $N$-body simulations of structure growth with physical
 modelling of galaxy evolution to investigate whether the shift in
 cosmological parameters between the 1-year and 3-year results from
 the Wilkinson Microwave Anisotropy Probe affects predictions for the
 galaxy population.  Structure formation is significantly delayed in
 the WMAP3 cosmology, because the initial matter fluctuation amplitude
 is lower on the relevant scales. The decrease in dark matter
 clustering strength is, however, almost entirely offset by an
 increase in halo bias, so predictions for galaxy clustering are
 barely altered.  In both cosmologies several combinations of physical
 parameters can reproduce observed, low-redshift galaxy properties;
 the star formation, supernova feedback and AGN feedback efficiencies
 can be played off against each other to give similar results.  Models
 which fit observed luminosity functions predict projected 2-point
 correlation functions which scatter by about 10-20 per cent on large
 scale and by larger factors on small scale, depending both on
 cosmology and on details of galaxy formation.
 Measurements of the pairwise velocity distribution
 prefer the WMAP1 cosmology, but careful treatment of the systematics
 is needed. Given current modelling uncertainties, it is not easy to
 distinguish the WMAP1 and WMAP3 cosmologies on the basis of
 low-redshift galaxy properties. Model predictions diverge more
 dramatically at high redshift. Better observational data at $z>2$
 will better constrain galaxy formation and perhaps also cosmological
 parameters.
\end{abstract}

\section{Introduction}
\label{sec:intro}

Our current understanding of the cosmic evolution is based heavily on
measurements of the Cosmic Microwave Background (CMB).  The discovery
of the CMB \citep{penzias65} led to general acceptance of the Hot Big
Bang theory, and further support came from the subsequent
demonstration of its near-perfect black-body spectrum
\citep{mather90}.  The detection and continuing refinement of measures
of angular structure in the CMB confirmed theoretical predictions for
the growth of structure in flat cosmologies dominated by non-baryonic
Cold Dark Matter (CDM)
\citep{smoot90,deBernardis00,deBernardis02}.  The power spectrum
of this structure encodes information about the values of the
cosmological parameters, although degeneracies prevent a precise
determination of all parameters from CMB data alone.  By including
constraints from other kinds of data, it becomes possible to constrain
many cosmological parameters quite tightly.  During the last decade, a
growing body of such measurements have ushered in a new era of
`precision cosmology'.

The first-year data from the Wilkinson Microwave Anisotropy Probe
(WMAP) \citep{spergel03} did much to establish $\Lambda$CDM, a flat
CDM model with a cosmological constant, as the standard model for
structure formation. In addition, they provided apparently precise
estimates for a number of cosmological parameters.  Two further years
of WMAP data have significantly refined these estimates, leading to
noticeable shifts in some of the best values \citep{spergel07}. The
most important differences are a reduction in optical depth to the
last scattering surface ($\tau$), a lower value for the amplitude of
matter fluctuations on $8\,h^{-1}\,{\rm Mpc}$ scale ($\sigma_8$), a
reduction of the scalar spectral index on primordial perturbations
($n$), and a lowering of the total matter density ($\Omega_m$).

Several studies prior to the latest WMAP results suggested a lower value for
$\sigma_8$.  These looked at evolution in the abundance of galaxy clusters
\citep{borgani01,schuecker03}, and at constraints from the abundance and
clustering of low-redshift galaxies in combination with observed cluster
mass-to-light ratios \citep{vandenbosch03,tinker05}.  
Parameter estimates from these
methods are partially degenerate in $\sigma_8$ and $\Omega_m$ and other
studies, notably of cosmic shear, have tended to give conflicting indications
\citep[e.g,][]{massey07, benjamin07, hoekstra06}. Recent
  studies on giant arc statistics \citep{guoliang06} also suggest that 
  low values of $\sigma_8$ may be difficult to
  reconcile with the observations. It is only the shrinking of the allowed
parameter region forced by the new WMAP data that has persuaded much of the
astronomical community to prefer a ``standard'' model with a lower value of
$\sigma_8$. It should be remembered, however, that including parameters beyond
the usual minimal set significantly relaxes constraints, so that both first-
and third-year WMAP parameter sets should probably be treated as plausible.

Modifications of the cosmological parameters of the kind discussed
above can significantly affect the formation and evolution of
structure.  Recent studies have discussed the implications of WMAP
third-year results for the formation of the first stars and black
holes, and for the reionization of the intergalactic medium
\citep{gao06,alvarez06,popa06,lewis06,iliev07}, finding significantly
later formation and reionization than previously thought.

The present paper investigates the impact of this delay on the observed
properties of galaxies, both at low and at high redshift.  We combine
high-resolution $N$-body simulations with semi-analytic modelling techniques
to simulate the evolution of the galaxy population
\citep{springel05a,croton06,delucia07}.  Our paper is organised as follows:
the $N$-body simulations and the semi-analytic modelling assumptions are
described in sections~\ref{sec:simulations} and~\ref{sec:sam}.  In
section~\ref{sec:dm} we discuss the formation of dark matter structures in
cosmologies with first-year and third-year WMAP parameters, while in
section~\ref{sec:samresults} we compare simulations of the evolution of the
galaxy population in these two cosmologies.  Finally, in
section~\ref{sec:conclusions} we summarise and discuss our findings. For the
convenience of the reader we also provide an appendix containing an analytic
comparison of evolution in the abundance and clustering of dark halos in our
two cosmologies, based on the formulae of \citet{mo02}

\section{The Simulations}
\label{sec:simulations}

We have carried out two simulations of the growth of structure in a
$\Lambda$CDM Universe using the cosmological parameter sets listed in
Tab.~\ref{tab:cosmparam}.  The cosmological parameters used for our
WMAP1 simulation are derived from a combination of first-year WMAP
results \citep{spergel03} with the 2dFGRS galaxy power spectrum
\citep{colless01}, and correspond to those used for the
\emph{Millennium Simulation} \citep[MS; ][]{springel05a}.  Our WMAP3
simulation adopts cosmological parameters derived from a combination
of third-year WMAP data on large scale, and Cosmic Background Imager
(CBI) and extended Very Small Array (VSA) data on small scale
\citep{spergel07}.  We note that, among the different sets of
cosmological parameters consistent with the third-year WMAP data and
other observational data, we have chosen to look at the one which
differs the most from the parameters of our WMAP1 simulation.

The most significant difference between the two sets of parameters
listed in Table~\ref{tab:cosmparam} is in the lower value of $\sigma_8$
adopted for the WMAP3 simulation.  Our WMAP1 and WMAP3 cosmologies
also differ in the scalar spectral index of primordial density
perturbations ($n = 1 \rightarrow 0.947$), and in the matter density
($\Omega_m = 0.25 \rightarrow 0.226$).  As noted above, a number of
recent studies have shown that this change in cosmology results in a
significant delay of structure formation in the WMAP3 case in
comparison to WMAP1. We will discuss this in more detail in the
following section.

The numerical parameters used for our simulations are listed in
Table~\ref{tab:numparam}. The mass and force resolution are the same as used
for the MS, while the volume is a factor $64$ smaller.  We have stored the data
at the same $64$ output times as in the MS. These are approximately
logarithmically spaced between $z=20$ and $z=1$ and approximately linearly
spaced in time thereafter.  Friends-of-friends (FOF) group catalogues were
computed on the fly for each snapshot, and the algorithm {\small SUBFIND}
\citep{springel01} was employed to decompose each FOF group into a set of
disjoint substructures.  As in \citet{springel05a}, only substructures which
retain at least $20$ bound particles after a gravitational unbinding procedure
are considered to be genuine substructures.  Substructure catalogues are then
used to construct detailed merger history trees that provide the basic input
needed for the semi-analytic model described in the next section.  We refer to
\citet{springel05a} for more details of the merger-tree construction.  Both our
simulations were run using the tree-based parallel code {\small GADGET2}
\citep{springel05b}.  The initial power spectra were generated using CMBFAST
\citep{Seljak96} with the cosmological parameters listed in Tab. 1 as input.
The Fourier modes of the initial density field in the two cases were identical
except for the amplitude adjustment needed to reproduce the correct power
spectrum. Thus structures correspond closely in the two cases.

\begin{table}
\caption{Cosmological parameters of the two simulations used in this
  study. $\Omega_m$, $\Omega_{\Lambda}$, $\Omega_b$ are the density of matter,
  dark energy and baryons respectively. $\sigma_8$ and $n$ are the amplitude of
  the mass density fluctuations, and slope of initial power spectrum. The
  Hubble constant is parameterised as $H_0 = 100\, h\, {\rm km\, s^{-1}
  Mpc^{-1}}$. } 
\begin{center}
\begin{tabular}{cccccccccc} \hline
$ $ & ${\rm WMAP1}$ & ${\rm WMAP3}$ \\ \hline
$\Omega_m$ & 0.25 & 0.226 \\
$\Omega_{\Lambda}$ & 0.75 & 0.774 \\
$\Omega_b$ &0.045 & 0.04 \\
$\sigma_8$& 0.9 & 0.722 \\
$h$ & 0.73   & 0.743 \\
$n$ & 1 & 0.947 \\
\hline
\end{tabular}
\end{center}
\label{tab:cosmparam}
\end{table}

\begin{table}
\caption{Numerical parameters of our two simulations. $L$ is the box size,
  $n_{\rm p}$ is the particle number, $\epsilon$ is the softening length, and
  $m_{\rm p}$ is the particle mass.}
\begin{center}
\begin{tabular}{cccccccccc} \hline
$L$ & $n_{\rm p}$ & $\epsilon$ & $m_{\rm p}({\rm WMAP1})$ & $m_{\rm p}({\rm WMAP3})$ \\ \hline
$125\,{\rm Mpc}\,h^{-1}$ & $540^3$ & $5{\rm kpc}$ & $8.61e8{\rm
  M_{\odot}}\,h^{-1}$ & $7.78e8 {\rm M_{\odot}}\,h^{-1}$\\ 
\hline
\end{tabular}
\end{center}
\label{tab:numparam}
\end{table}

\begin{figure}
\bc
\hspace{-0.1cm}
\resizebox{8cm}{!}{\includegraphics{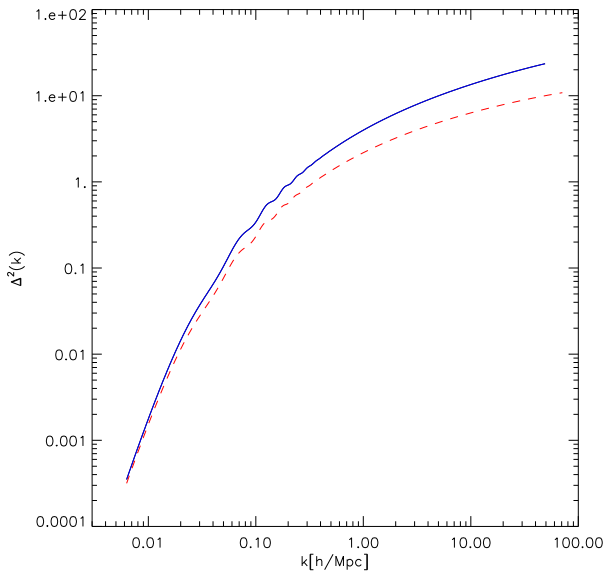}}\\%
\caption{The power spectra of our two simulations: WMAP1 (blue solid) and
WMAP3(red dashed). These were generated using CMBFAST with the cosmological
parameters listed in Tab.~\ref{tab:cosmparam}. The quantity plotted is
$\Delta^2(k) \propto k^3P(k)$. }
\label{fig:ps}
\ec
\end{figure}

Fig.~\ref{fig:ps} shows the power per decade in the linear initial
conditions of our WMAP1 (blue) and WMAP3 (red) simulations.  The
significantly lower value of $\sigma_8$ in the WMAP3 case reduces the
overall normalisation.  In addition, the red tilt in the primordial
power spectrum index makes the difference greater on small scales than
on large scales. Indeed, on the scales responsible for the low $\ell$
microwave background fluctuations the two power spectra are almost the
same, reflecting the fact that they are both required to fit observed
CMB fluctuations on these scales. Conversely, the suppression of power
is particularly significant on the small scales responsible for the
formation of the first nonlinear structures.

\section{The Semi-Analytic Model}
\label{sec:sam}

In this paper we use the galaxy formation model described in
\citet{delucia07} which is based on the WMAP1 cosmology and builds on
previous work by the ``Munich'' galaxy formation group
\citep{kauffmann99a,springel01,delucia04,springel05a,croton06}.
Although not in perfect agreement with all aspects of the local galaxy
population (see, for example, \citet{weinmann06}), this model does
quite a good job of reproducing the observed relations between stellar
mass, gas mass, and metallicity \citep{delucia04}, the observed
luminosity, colour, and morphology distributions
\citep{croton06,delucia06} and the observed two-point correlation
functions \citep{springel05a}.  \citet{kitzbichler07} have recently
shown that it also agrees reasonably well with the observed galaxy
luminosity and mass functions at high redshift.  We refer the reader
to the original papers for a full description of the numerical
implementation, and of the physical processes modelled.  In the
following, we summarise briefly the treatment of those physical
processes for which we needed to change the efficiency parameters in
order to maintain agreement with observations of the local galaxy
population when we switch from WMAP1 to WMAP3 (see
Table~\ref{tab:sam} and Sec.~\ref{sec:samresults}).

In the semi-analytic model we use in this work, star formation is
assumed to occur at a rate given by:
\begin{equation}
\dot{m}_{*}=\alpha_{\rm SF}(m_{\rm cold}- m_{\rm crit})/t_{\rm dyn,disc}
\end{equation}
where $m_{\rm cold}$ and $t_{\rm dyn,disc}$ are the cold gas mass and the
dynamical time of the galaxy, defined as the ratio between the disk radius and
the virial velocity, $m_{\rm crit}$ corresponds to a critical value for the gas
surface density \citep{Kennicutt98,Kauffmann96,mo98}, and $\alpha_{\rm SF}$
controls the efficiency of the transformation of cold gas into stars when the
gas surface density is above the critical value. (See \citet{croton06}
for more detailed descriptions of the implementation of this process
and of the feedback processes described below.)

Massive stars explode as supernovae shortly after a star formation event and
are assumed to reheat a gas mass that is proportional to the mass of stars
formed \citep[based on the observations of][]{martin99}:
\begin{equation}
{\Delta}m_{\rm reheated} = \epsilon_{\rm disk}{\Delta}m_{*},
\end{equation}
Again following \citet{croton06}, we write the energy released by an
event which forms a mass ${\Delta}m_{*}$ in stars as:
\begin{equation}
{\Delta}E_{\rm SN}=0.5\epsilon_{\rm halo}{\Delta}m_{*}V^2_{SN},
\end{equation}
where $0.5V^2_{\rm SN}$ is the mean supernova energy injected per unit
mass of newly formed stars, and $\epsilon_{\rm halo}$ represents the
efficiency with which this energy is able to convert cold interstellar
medium into hot, diffuse halo gas. The amount of gas that leaves the
dark matter halo in a ``super-wind'' is determined by computing
whether excess SN energy is available to drive the flow after
reheating of material to the halo virial temperature.

As in \citet{kh00}, black holes are formed and fuelled during mergers:
\begin{equation}
{\Delta}m_{\rm BH} = \frac{f'_{\rm BH} m_{\rm cold}}{1+(280 {\rm km}{\rm
    s}^{-1}/V_{\rm vir})^2}
\end{equation}
We assume here that black holes grow during both major and minor mergers, and
that the efficiency of gas accretion onto the black hole scales with the
baryonic mass ratio of the merging galaxies:
\begin{equation}
  f'_{\rm BH} = f_{\rm BH}\times(m_{\rm sat}/m_{\rm central})
\end{equation}
This is the primary process driving the growth of the total mass in
supermassive black holes.  Individual black holes can also gain mass
through merging when their host galaxies merge.

Finally, we use the model by \citet{croton06} to describe heating by
centrally located AGN in massive groups and clusters. This process is
assumed to be associated with ``radio mode'' outflows which suppress
cooling flows and thus the condensation of gas onto the central
galaxies.  The process is assumed to occur whenever a massive black
hole finds itself at the centre of a static hot gas halo.  During this
phase, the accretion rate onto the central supermassive black hole is
taken to be:
\begin{equation}
\dot{m}_{\rm BH,R}=\kappa_{\rm AGN}
\left(\frac{m_{\rm BH}}{10^8M_{\odot}}\right)
\left(\frac{f_{\rm hot}}{0.1}\right)
\left(\frac{V_{\rm vir}}{200 \rm {kms^{-1}}}\right)^3,
\end{equation}
where $m_{\rm BH}$ is the black hole mass, $f_{\rm hot}$ is the fraction of the
total halo mass in the form of hot gas, $V_{\rm vir}$ is the virial velocity of
the halo, and $\kappa_{\rm AGN}$ is efficiency parameter with units of
$M_{\odot}\rm yr^{-1}$. The energy released during this accretion process is
used to reduce the cooling flow.  \citet{croton06} showed that this results
in complete suppression of cooling in relatively massive haloes and groups.  The
process starts being effective at a mass scale that evolves as a function of
redshift as shown in their Fig.~7.

In our simulation scheme, haloes (and the galaxies within them) are
followed even after they are accreted onto larger systems.  The
dynamics of such a satellite subhalo is followed explicitly by the
N-body simulation until tidal stripping causes its mass to fall below
the resolution limit of the simulation
\citep{ghigna00,delucia04a,gao04}.  When this happens, we estimate a
survival time ($t_{\rm merge}$) for the associated galaxy using its
current orbit and the classical dynamical friction formula of
\citet{bt87}.  Once this time has elapsed, the galaxy is assumed to
merge onto the central galaxy of its current halo. While it still
survives it is assumed to follow the particle which was the most bound
particle of the subhalo at the last time it was identified.
\citet{delucia07} found that increasing the merging time by a factor
of $2$ slightly improves the fit to observed luminosity function.
Such an increase has other effects which \citet{delucia07} did not
study, for example, it increases the amplitude of small-scale galaxy
correlations by about a factor of 2 at $r < 100$~kpc. Others authors
have claimed that this merging time should be effectively set to zero,
so that such ``orphan'' galaxies lose their identity at the same time
as their subhaloes \citep{conroy07}. Here we leave this issue for
detailed study in future work and simply consider the pre-factor to be
used for $t_{\rm merge}$ as a free parameter.

\section{Formation of dark matter structures}
\label{sec:dm}

\begin{figure*}
\bc
\resizebox{18cm}{!}{\includegraphics{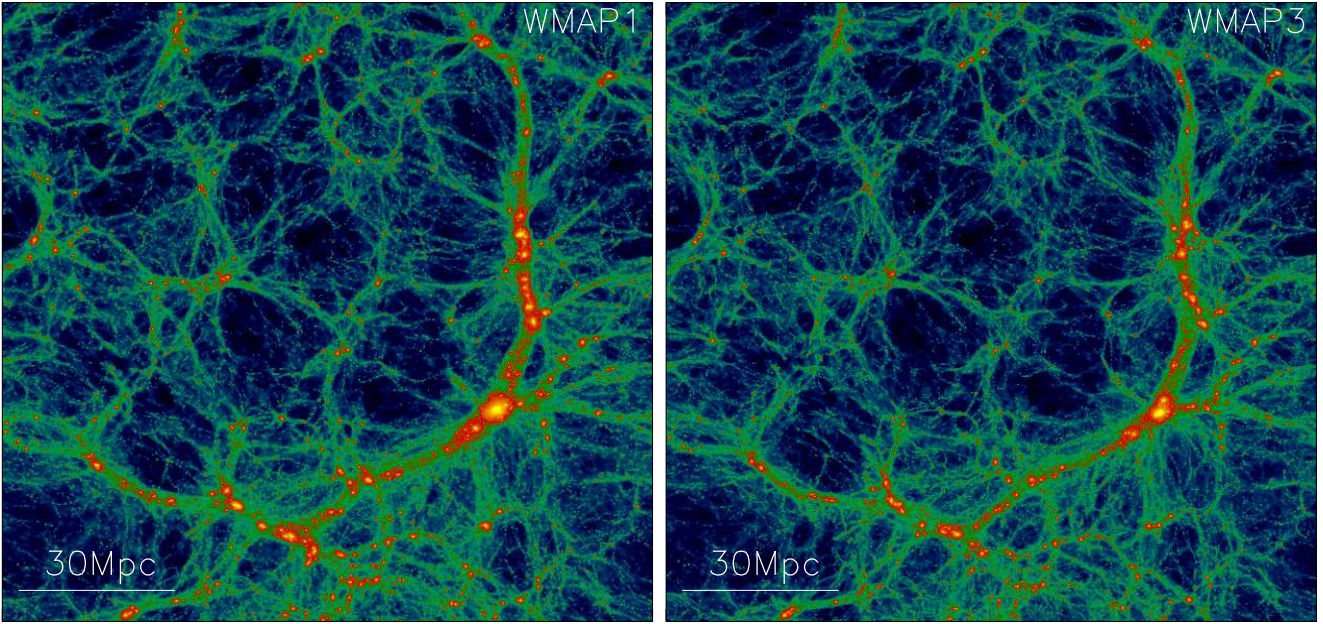}}\\%
\caption{Redshift zero distribution of dark matter within a slice of thickness
  $20\,{\rm Mpc}\,{h}^{-1}$ through our two simulations, WMAP1 (left) and
  WMAP3 (right).}
\label{fig:twobox}
\ec
\end{figure*}
Figure~\ref{fig:twobox} shows the dark matter distribution at $z=0$
within a slice $20\,{\rm Mpc}\,{h}^{-1}$ thick cut from the full
volume of our simulations.  The WMAP1 model is on the left and the
WMAP3 model on the right.  The projection is colour-coded by density
and provides a visual illustration of the delay of structure
formation in the lower $\sigma_8$ model. Although the overall
structure is very similar, it is clear that massive haloes lying
at the nodes of the cosmic web are in a more advanced stage of
merging in the WMAP1 case than in the WMAP3 case.

\begin{figure}
\bc
\resizebox{8cm}{!}{\includegraphics{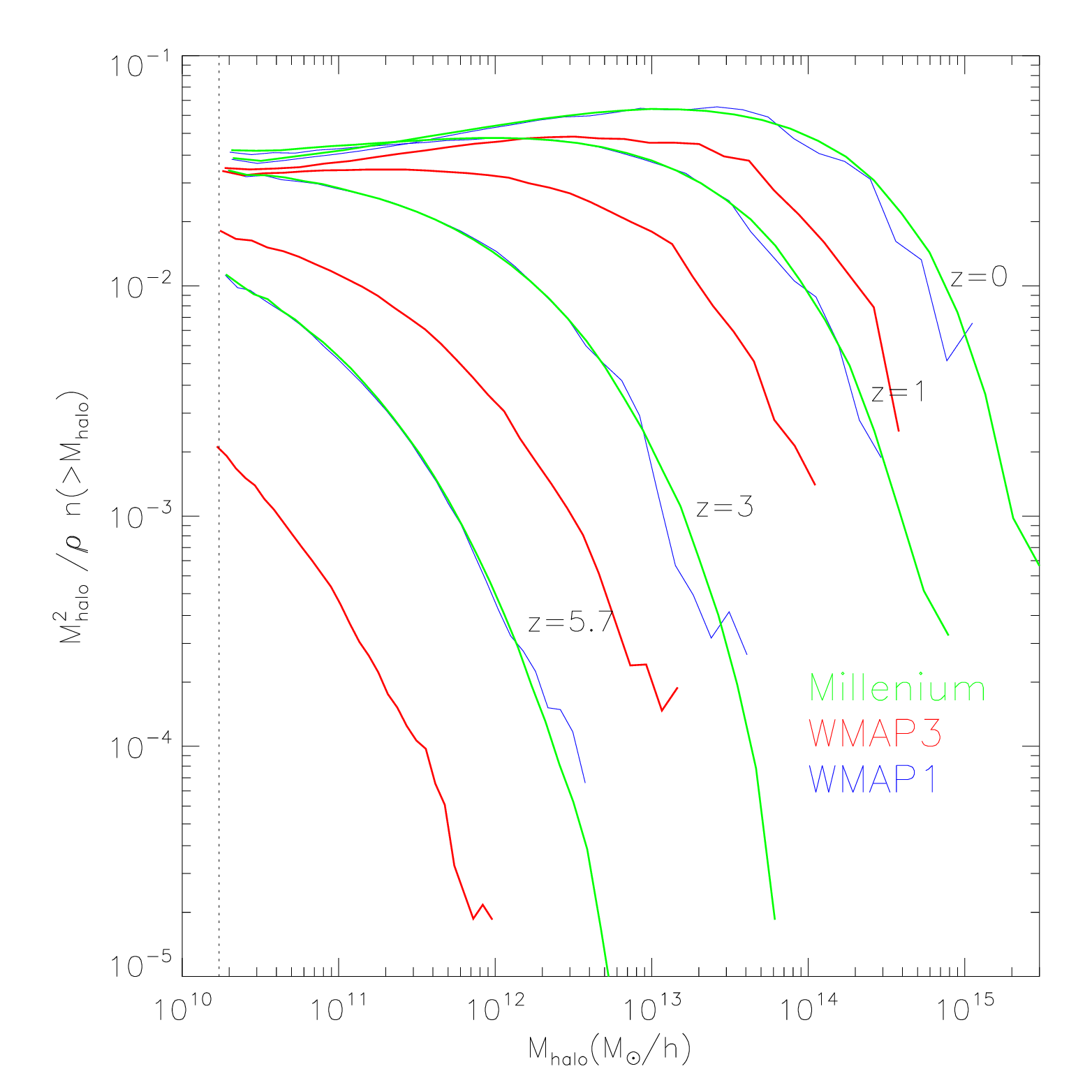}}\\%
\caption{Cumulative halo number density as a function of halo mass and
redshift. Results for the WMAP3 and WMAP1 simulations are represented
by red and blue curves respectively.  Results for the Millennium
Simulation are shown in green. The halo abundance is multiplied by
$M^2$ in order to reduce the dynamic range of the ordinate.}
\label{fig:mf}
\ec
\end{figure}

The differing fluctuation amplitudes of the two simulations translate
into a systematic difference between their halo mass functions.  This
is illustrated in Fig.~\ref{fig:mf} where we compare the cumulative
number density of halos for the two cases (blue for WMAP1, red for
WMAP3). In addition, we show the corresponding function for the much
larger Millennium Simulation (green) which used our WMAP1 cosmological
parameters. It agrees very well with the smaller WMAP1 simulation of
this paper.  At $z = 0$, the most massive haloes in our WMAP3
simulation are about 1.5 times less massive than their counterparts in
the WMAP1 simulation.  The number of haloes more massive than
$10^{13}\,{\rm M}_{\odot}\,h^{-1}$ is $\sim 25$ per cent smaller in
the WMAP3 case than in the WMAP1 case.  These differences increase at
higher redshift. At $z\sim6$ (i.e. at the end of the reionization
epoch) the most massive haloes in the two simulations differ by about
a factor of 5 in mass.

We recall that the main differences between the cosmological
parameters used in the two simulations are the lower value of
$\sigma_8$ and the redder primordial power spectrum index $n$ in the
WMAP3 case.  As shown in Fig.~\ref{fig:ps}, these combine to produce a
substantial suppression of small-scale power.  This has important
implications for the formation of the first objects and for the star
formation history at early times. \citet{gao06} studied this problem
using high-resolution cosmological simulations with a smoothed
particle hydrodynamics treatment of baryonic processes.  They found a
much lower abundance of potentially star-forming halos at high
redshift for WMAP3 than for WMAP1.  This reduction in the number of
`mini-haloes' at $z\sim 10$ was also discussed by \citet{reed07} using
analytic models to explore the dependence on cosmological
parameters. For fluctuation amplitudes at the WMAP1 level, very
efficient production of UV radiation is needed to reionize the
intergalactic medium by $z\sim 15$, as required by the WMAP1 value for
the electron scattering optical depth $\tau$ \citep[e.g.][]{ciardi03}.
Interestingly, the delay in structure formation in a WMAP3 Universe is
such that {\it equally} efficient UV production is needed to reionize
by $z\sim 10$, as inferred from the WMAP3 value of $\tau$
\citep{alvarez06}.

\begin{figure}
\bc
\resizebox{8cm}{!}{\includegraphics{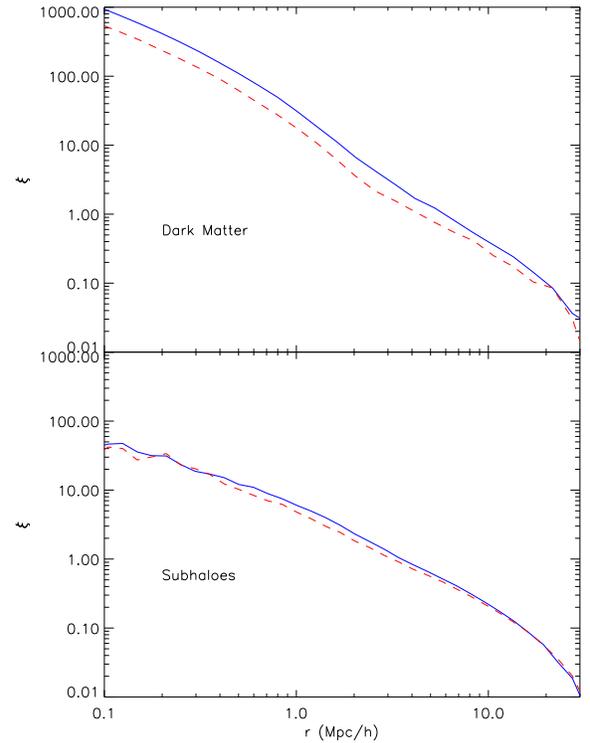}}\\%
\caption{Autocorrelation functions at $z=0$ for dark matter (top
panel) and for all subhaloes with at least 20 particles (bottom
panel). Blue solid and red dashed curves show results for our WMAP1 
and WMAP3 simulations respectively. }
\label{fig:cf_dm}
\ec
\end{figure}

Fig.~\ref{fig:cf_dm} shows $z=0$ autocorrelation functions in our
WMAP1 and WMAP3 simulations for the dark matter (in the top panel) and
for all dark matter substructures identified by {\small SUBFIND} (in
the bottom panel; this corresponds to all self-bound (sub)haloes with
more than 20 simulation particles).  The difference in fluctuation
amplitude causes the correlation strength to be almost a factor of two
smaller in the WMAP3 case than in the WMAP1 case on scales below a few
Mpc. Curiously, however, this decrease is almost completely
compensated by an increase in the (sub)halo bias, so that the 2-point
correlation functions for subhaloes are almost identical in the two
cases. There is a slight residual offset on scales $0.5\,{\rm
Mpc}h^{-1} \lesssim r \lesssim 5\,{\rm Mpc}h^{-1}$. As we will see
below, this results in very similar galaxy correlation functions being
predicted in the two cases.

\section{Galaxy Formation}
\label{sec:samresults}

In this section we analyse galaxy populations 
simulated using the semi-analytic framework presented in
Sec.~\ref{sec:sam}.  Table~\ref{tab:sam} lists the combinations of
semi-analytic parameters for which we will show detailed results.
Model A is the parameter set used by \citet{delucia07} for the
Millennium Simulation. As expected, this also gives good agreement
with observation for our WMAP1 simulation.  Models B and C are
parameter combinations that, as we show below, produce a similar
agreement with local data for the WMAP3 cosmology.

\begin{table}
\caption{Principal parameters of our galaxy formation models (see text for
  details).} 
\begin{center}
\begin{tabular}{cccccccccc} \hline
$$ & $A$ & $B $ & $C$ \\ \hline
$\alpha_{\rm SF}$ & 0.03 & 0.03 & 0.07\\
$\epsilon_{\rm halo}$ & 0.35&0.2 & 0.28\\
$\epsilon_{\rm disk}$& 3.5&3 & 4.5\\
$\kappa_{\rm AGN}$ & 7.5e-6&6.5e-6 &1.2e-5 \\
$f_{\rm BH}$ &0.03 &0.03 &0.05 \\
$t_{\rm merge}$ & $2 \times t_{\rm friction}$ & $ t_{\rm friction}$ &
$2 \times t_{\rm friction}$\\
${\rm Best~for}$ & $\rm WMAP1$ & $\rm WMAP3$ & $\rm WMAP3$\\
\hline
\end{tabular}
\end{center}
\label{tab:sam}
\end{table}

Model B has the same star formation efficiency as model A
($\alpha_{\rm SF}$), but lower supernova feedback efficiency
($\epsilon_{\rm halo}$ and $\epsilon_{\rm disk}$), and lower AGN
feedback efficiency ($\kappa_{\rm AGN}$). For this model, we have also
eliminated the pre-factor of $2$ which \citet{delucia07} introduced in
the definition of the merging time.  In contrast, in model C, we
double the star formation efficiency (relative to model A) in order to
compensate for the delay in structure formation, but this must be
compensated by much higher feedback efficiencies (both from supernovae
and from AGN) to prevent the overproduction of stars at late times.
For this model we keep the pre-factor of $2$ in the definition of the
merging times of satellite galaxies. We also increase the efficiency
of accretion onto black holes during mergers (which increases the
effectiveness of the `radio' mode - see Eq.~6).  In the following, we
will show that models B and C give similar $z=0$ results for the WMAP3
cosmology, and that these resemble the results of model A for the
WMAP1 simulation.  This shows that there are at least two independent
way to `compensate' for the delay in the structure formation within
this physical framework: we can decrease feedback on all scales while
keeping the same star formation efficiency (model B), or we can
increase both the star formation efficiency and the feedback
efficiency on all scales (model C).

In the following sections we discuss how the predictions of our three
models compare to observational data both in the local Universe and at
high redshift.

\subsection{Low-redshift Luminosity Functions}

\begin{figure}
\bc
\resizebox{8cm}{!}{\includegraphics{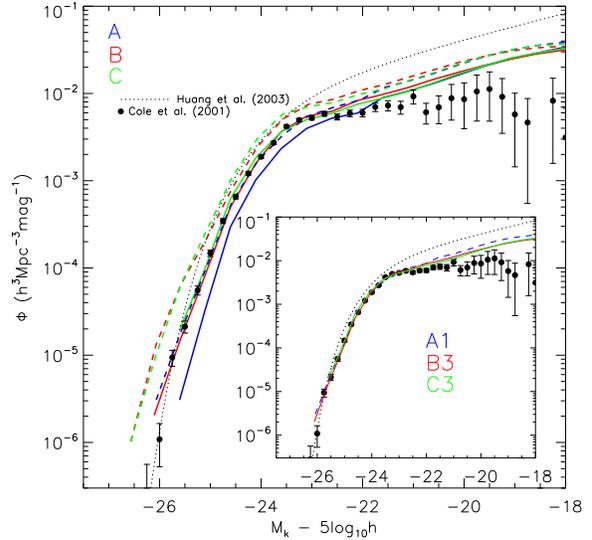}}\\%
\caption{K-band luminosity function for the three galaxy formation
models and the two cosmologies simulated in this study.  Results from
models A, B, and C are shown in blue, red and green respectively.
Solid and dashed lines correspond to our WMAP3 and WMAP1 simulations
respectively.  The black symbols with error bars show the
observational determination by \citet{cole01}, while the black dotted
curve shows the measurements of \citet{huang03}.  The inset repeats
the figure but shows results only from model A applied to WMAP1 and
from models B and C applied to WMAP3.}
\label{fig:lf_k2}
\ec
\end{figure}

\begin{figure}
\bc
\resizebox{8cm}{!}{\includegraphics{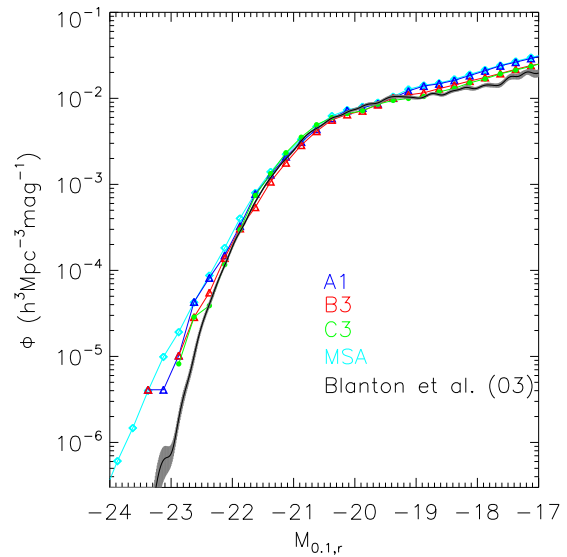}}\\%
\caption{Galaxy luminosity function at SDSS $r_{0.1}$ band. Coloured
lines show predictions from the three ``good'' models of this paper.
We also plot results from model A applied to the Millennium
Simulation. The black curve with an error band corresponds to the
observational measurement from \citet{blanton03b}.}
\label{fig:lf_r}
\ec
\end{figure}

Figure~\ref{fig:lf_k2} compares observational estimates of the K-band
luminosity function of nearby galaxies to predictions from our three
galaxy formation models (differentiated by colour) applied to
each of our two simulations (differentiated by line type). In both
cosmologies models B and C give almost identical results, while model
A predicts fewer galaxies at bright luminosities ($ M_{\rm K} < -22$)
by a factor that varies from $1.3$ to $3$.  The inset in
Fig.~\ref{fig:lf_k2} repeats the figure showing only results from
model A applied to our WMAP1 simulation and from models B and C
applied to our WMAP3 simulation. This shows all three models to
produce similarly good fits to the observations around and above the
`knee'.  All three models show an excess of galaxies fainter than
${\rm M}_{\rm K} - 5\,{\rm log} h \sim -22$ with respect to the
\citet{cole01} data, although they lie below the luminosity function
given by \citet{huang03}.  This possible excess is more pronounced in
model A. In the following we will only discuss results from our
`best' models (i.e. model A for our WMAP1 simulation and for the
Millennium Simulation, models B and C for our WMAP3 simulation; in the
following we denote these as A1, MSA, B3 and C3).

Figure~\ref{fig:lf_r} compares model galaxy luminosity functions in the Sloan
Digital Sky Survey (SDSS) r-band to observational measurements from the SDSS
itself. Here the observational uncertainties are much smaller than for the
K-band luminosity function of Fig.~\ref{fig:lf_k2}.  Since the observed
absolute magnitudes quoted by \citet{blanton03a} are band-shifted to $z=0.1$,
we also correct our simulated r-band absolute magnitude to this redshift using
their K-correction code.  Fig.~\ref{fig:lf_r} shows all our ``good'' models to
agree very well with the observational data, particularly around the knee of
the luminosity function ($-22 < M_r < -19$). All models overpredict the
abundance of faint galaxies, although the problem appears significantly less
dramatic here than in Fig.~~\ref{fig:lf_k2}. In all models the rarest and most
luminous galaxies are also too bright by 0.2 to 0.6 magnitudes.  This problem
may be ,at least partially, due to our assigning light to these objects which
should be part of the intracluster light of the groups or clusters of which
they are the central galaxies (see also \citet{conroy07}).

\subsection{The Tully-Fisher Relation}
\label{sec:TF}

\begin{figure}
\bc
\resizebox{8cm}{!}{\includegraphics{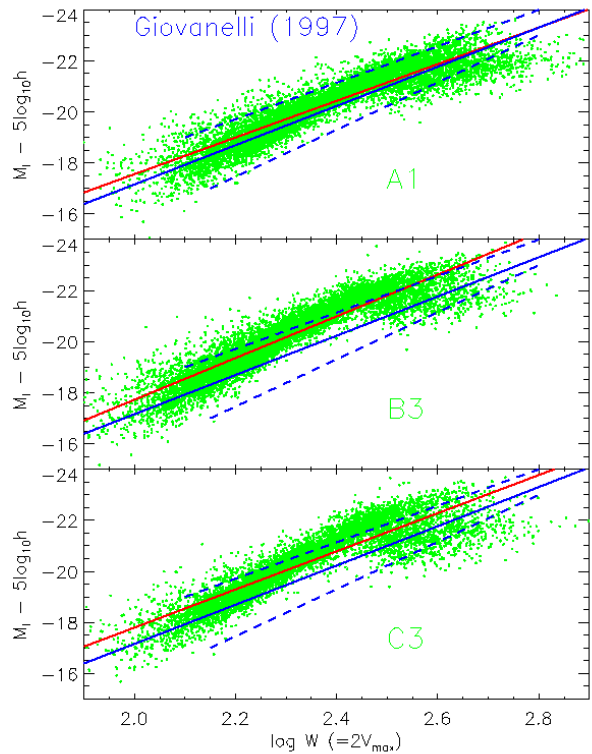}}\\%
\caption{Tully-Fisher relation for spiral galaxies in our three
models. We show only central galaxies with $1.5< M_{B,\rm
bulge}-M_{B,\rm total}<2.6$, and we approximate the observed HI
linewidth by twice $V_{\rm max}$ for the dark halo, as measured
directly in our simulations. Red lines are linear fits to the model
results. The blue solid line shows the mean observational relation by
\citet{giovanelli97}, with blue dashed lines indicating the scatter in
this relation.}
\label{fig:tf}
\ec
\end{figure}

In Fig.~\ref{fig:tf}, we compare the Tully-Fisher relation for our
model galaxies to the observational determination by
\citet{giovanelli97}. Green symbols show all model central galaxies
with $1.5< M_{B,\rm bulge}-M_{B,\rm total}<2.6$ where $M_{B,\rm
bulge}$ and $M_{B,\rm total}$ are the absolute magnitude of the bulge
and the total B-band magnitude, respectively.  This selection is made
in order to isolate Sb/c spirals as in the observed sample of
\citet{giovanelli97}. The mean observational relation is shown by a
solid blue line in the figure, with the corresponding $1 \sigma$
scatter indicated by the dashed lines.  The relation of Giovanelli et
al. is already corrected for internal extinction.  We
therefore do not include dust effects when predicting I--band
magnitudes for this plot.  Red lines in each panel show a linear fit
to the model results.  The top panel of Fig.~\ref{fig:tf} demonstrates
that model A1 reproduces both the slope and the zero-point of the
observed relation \citep[as also shown in][]{croton06}.  Models B3 and
C3 exhibit a substantially brighter zero-point than observed.  This
occurs because fitting the observed luminosity function within the
less evolved halo mass function of the WMAP3 cosmology (see Fig.~2)
requires us to put galaxies of given luminosity at the centre of lower
mass halos than in the WMAP1 case.

Finding theoretical models which simultaneously fit both the observed
luminosity functions and the observed Tully-Fisher relation has always been
difficult\citep[e.g.][]{cole00}.  The results in Fig.~\ref{fig:tf} show that
this remains true. Note, however, that it is uncertain how best to extract
quantities from our models for comparison with the observational data.  For
example, we assume here that $W$ (the measured HI linewidth) is twice the
maximum circular velocity of the galaxy's dark halo, as measured directly from
the simulation and without any correction for the effects of the baryonic mass
of the galaxy.  This is a rough assumption which may be systematically in
error. In particular, if galaxy formation results in a maximum disk rotation
velocity which is significantly and systematically above this value, the
population of galaxies in our WMAP3 model could be reconciled with the
observations. Changing the transformation between these two velocities can
change the slope and zero-point, or introduce curvature into the model
Tully-Fisher relation. Semi-analytic models provide robust estimates for the
locations, velocities and global properties (luminosity, colour, stellar mass,
etc,) of galaxies, but become less reliable when quantities (such as $W$)
which depend sensitively on the internal structure of the galaxies must be
predicted \citep[see also the discussion in][]{somerville99}.

\subsection{Mass-to-Light Ratios of Clusters}
\label{subsec:masslum}
\begin{figure}
\bc
\hspace{-0.9cm}
\resizebox{9cm}{!}{\includegraphics{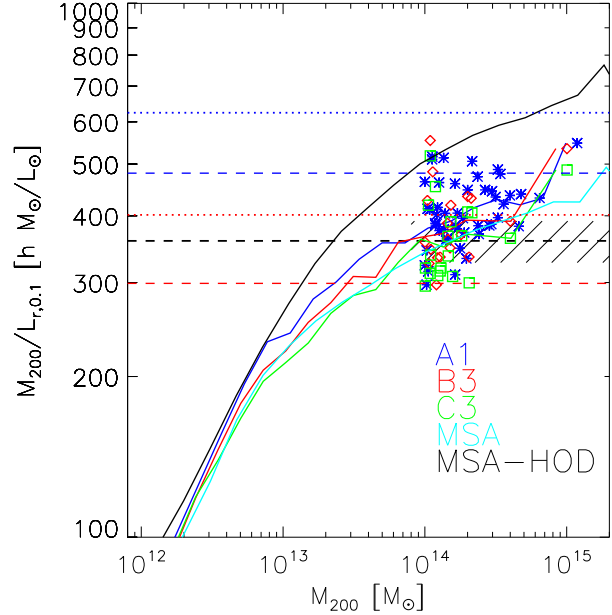}}\\%
\caption{Mass-to-light ratios in the SDSS $r_{0.1}$
band as a function of halo mass. Symbols show results for the 48 (20)
clusters with mass $M > 10^{14} M_\odot$ in our WMAP1 (WMAP3)
simulation. Solid lines show a running median over a wider range of
halo masses.  Blue, red, green and cyan are used for models A1, B3, C3
and MSA respectively. The black solid curve refers to the HOD model of
\citet{wang06} applied to the Millennium Simulation (MSA-HOD). The
black dashed line and the hatched area show the region occupied by the
observational data of \citet{carlberg96}. The horizontal dotted and
dashed lines show predictions for M/L of clusters 
$M>10^{14} M_{\odot}$ from \citet{vandenbosch03} and
\citet{tinker05} respectively. In each case blue refers to the WMAP1
and red to the WMAP3 prediction.}
\label{fig:m_l}
\ec
\end{figure}

In Fig.~\ref{fig:m_l}, we plot SDSS $r_{0.1}$ band mass-to-light
ratios ($M/L_{r,0.1}$) for individual clusters with $M \geq
10^{14}M_{\odot}$ (symbols) and the running median value over the full
resolved mass range (solid lines). Both $M$ and $L_{r,0.1}$ are
computed within $R_{200}$, defined as the radius within which the mean
mass overdensity is 200 times the critical value. Blue, red and green
are used for models A1, B3, and C3 respectively. Cyan is used for
the model MSA which provides a larger number of clusters than the
smaller simulations used in our study. The black horizontal line and
hatched area show the region occupied by the observational data in
\citet[][see Tinker et al. 2005 for details on the conversion to the
SDSS $r_{0.1}$ band]{carlberg96}. Our models exhibit a very weak mass
dependence over the observed mass range and agree reasonably well
with the observational measurements. The differences between the two
cosmological models, and between the two galaxy formation models for
same WMAP3 cosmology are small, even at small masses where
\citet{vandenbosch07} found a more pronounced decrease in the
mass-to-light ratio for WMAP3 cosmology with respect to WMAP1.

It is interesting to compare our results with those based on halo
occupation distribution (HOD) models. \citet{vandenbosch03} found that
their conditional luminosity function (CLF) model was unable to fit
the observed mass-to-light ratios of clusters in a cosmology with
WMAP1 parameters, and they argued strongly in favour of a cosmology
with a lower value of $\sigma_8$, similar to that subsequently
preferred by WMAP3. Once their models are constrained to fit the
observed luminosity and correlation functions of galaxies, they find
$\langle M/L\rangle$ to vary roughly as $\sigma_8^2$ on cluster
scales. \citet{tinker05} found a similarly strong dependence of the
mass-to-light ratio on $\sigma_8$ using an HOD model which differed in
detail and which they constrained with different observational data.
$M/L$ predictions from these models are shown in Fig.~\ref{fig:m_l} by
the horizontal dashed and dotted lines (blue and red are used for
WMAP1 and WMAP3 respectively). Predictions from \citet{vandenbosch03}
were obtained by converting their B-band estimates to the SDSS
$r_{0.1}$ band using the conversion factor adopted in
\citet{tinker05}.  These authors convert the mean B-band mass-to-light
ratio $<M/L_B>=363 h(M/L)_{\odot}$ of \citet{carlberg96} to
$<M/L_{r,0.1}>=359 h (M/L)_{\odot}$ in the SDSS $r_{0.1}$ band. Thus
we multiply the results of \citet{vandenbosch03} by a factor of
$359/363=0.988$. Fig.~\ref{fig:m_l} shows that the mass-to-light ratio
dependence on $\sigma_8$ in our models is much weaker than in the two
HOD models mentioned above. This is surprising since our models are
also a good fit to the observed galaxy luminosity functions, and fit
observed correlation functions moderately well, at least on large
scale (see below).

In Fig.~\ref{fig:m_l}, we also show as a black solid line the average
mass-to-light ratio predicted by the physically based HOD model
presented in \citet{wang06}. These authors use the same positions and
velocities for galaxies as our semi-analytic model (taken directly
from the Millennium Simulation) but adopt simple parametrized
functions to relate the star formation rates of galaxies to the mass
and location (satellite or central) of their parent subhalo. The
parameters of these functions are then minimized in order to optimize
the fit to the observed luminosity function and luminosity-dependent
correlation functions. The cluster mass-to-light ratios predicted by
this model are higher than those in any of the semi-analytic models in
our study, and are reasonably close to the predictions of
\citet{vandenbosch03} and \citet{tinker05} for a WMAP1 cosmology (the
cosmology adopted in the Millennium Simulation). Comparing the model
of \citet{wang06} in detail to the MSA model, we find that the larger
mass-to-light ratio on cluster scales is due to systematic lower
central galaxy luminosities (by a factor $\sim 2$) in the HOD
model. We note that \citet{delucia07} found reasonable agreement when
comparing this same MSA model to the observed K-band magnitudes of
brightest cluster galaxies (BCGs). We note also that precise
measurements of BCGs luminosities are complicated by the intrinsic
difficulty in separating the contribution of the cD envelope
\citep{Schombert88,Gonzalez05,Zibetti05}. The three HOD models shown
in Fig.~\ref{fig:m_l} exhibit a stronger dependence on cosmology (in
particular $\sigma_8$), but also a larger dependence on modelling
details than do our semi-analytic models. It seems, therefore, that a
good understanding of the consequences of modelling assumptions is
needed before firm conclusions can be drawn about $\sigma_8$ or other
cosmological parameters.

\begin{figure}
\bc
\resizebox{9.cm}{!}{\includegraphics{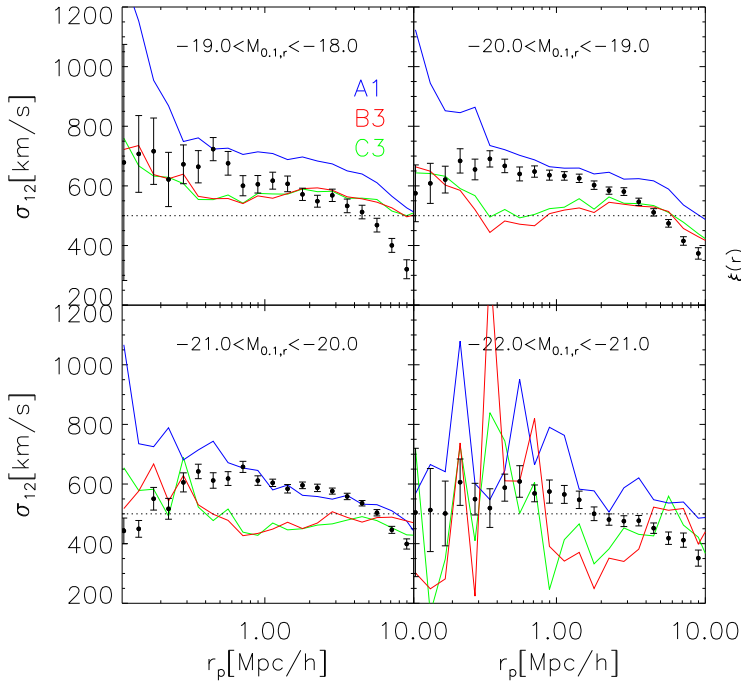}}\\%
\caption{Pairwise velocity dispersions for our three
SA models and for four luminosity bins in the SDSS $r_{0.1}$
band. Results from our three models are presented in blue (A1), red
(B3) and green (C3). The black symbols with error bars are
observational data from \citep{li07}. The horizonal dotted line
corresponds to the $\rm \sigma_{12}=500 km/s$}.
\label{fig:pvd}
\ec
\end{figure}

\subsection{Pairwise Velocity Dispersion}
Pairwise velocity dispersions (PVD) are a useful tool because of their
strong sensitivity to the abundance of massive haloes
\citep{mo93,jing04,yang04,tinker06}.  Fig.~\ref{fig:mf} shows that the
present-day abundance of massive haloes with mass $\geq 10^{14}
M_{\odot}/h$ in WMAP3 is almost $2.5$ times lower than the
corresponding abundance in WMAP1. This should be reflected in 
differences in the predicted PVD. 

In Fig.~\ref{fig:pvd}, we present the dependence of PVD on SDSS
$r_{0.1}$ luminosity for the three SA models used in this work. The
black points with error bars show observational estimates from
\citet{li07} obtained by modelling the full two-dimensional
redshift-space correlations. In order to compare our model prediction
with these data, we adopt the assumptions and modelling methods of
\citet{jing98,li07} except that rather than constructing large mock
catalogues, we measure directly both the 2-dimensional correlation
function in redshift space ($\xi(r_p,\pi)$) and the real space
correlation ($\xi(r)$) using the distant observer approximation for
the former. Fig.~\ref{fig:pvd} shows that model A1 produces higher
dispersions (by $\sim 50 - 150 \rm km/s$) on all scales and for all
luminosity subsamples than do the two WMAP3 models. Except for the
faintest galaxies, model A1 also gives a better fit to the
observational data. This agrees with \citet{li07} who compared their
measurements to the model of \citet{croton06} which is very similar to
our A1. Both WMAP3 models predict lower PVD than are observed, and
they also seem to show a different slope. Thus, PVD observations seem
to prefer a WMAP1 cosmology, as can be inferred from \citet{jing04}.

\subsection {Galaxy Clustering}
\label{sec:clustering}

\begin{figure}
\bc
\resizebox{9.cm}{!}{\includegraphics{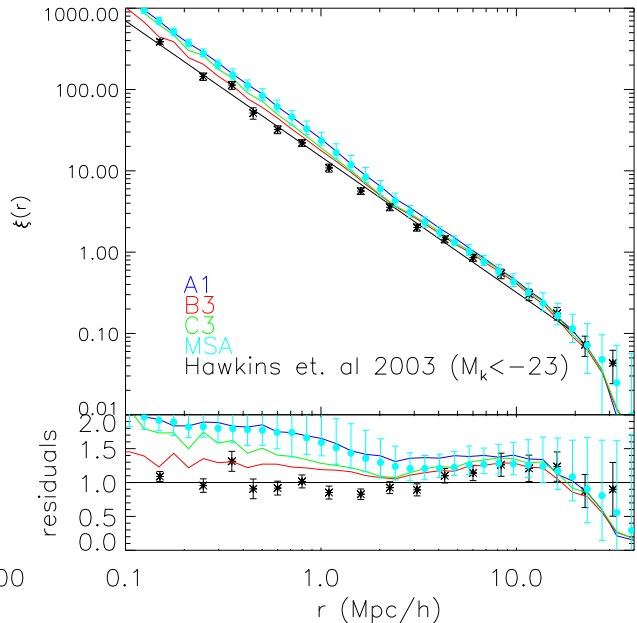}}\\%
\caption{2-point autocorrelation functions for luminous galaxies in
K-band at $z=0$. Solid colored curves show results for the three
``good'' models of this paper. The cyan symbols with error bars are
predictions based on 64 small boxes (each with the volume of the
simulations of this paper) cut from the Millennium Simulation
catalogue of \citet{delucia07}. Black symbols with error bars and
the black solid line show observational estimates and a power law fit
from the 2DFGRS\citep{hawkins03}. The residuals shown in the lower
panel are calculated with respect to this power law fit.}
\label{fig:cf_k}
\ec
\end{figure}

Figure~\ref{fig:cf_k} compares the 2-point correlation function of our
model galaxies at redshift $z = 0$ to a recent measurement from the
2dFGRS \citep{hawkins03}.  Solid colored lines show predictions from
our three ``good'' models (blue for model A1, red and green for models
B3 and C3 respectively).  For each model there are about 17000
galaxies with $M_{\rm k}<-23$.  This magnitude limit
is just fainter than the knee in Fig.~\ref{fig:lf_k2} so that most
luminous galaxies are included. In order to estimate the cosmic
variance in these estimates, we divide the \emph{Millennium
Simulation} galaxy catalogue of \citet{delucia07} into 64 small boxes,
each with the same volume as our new simulations, and then calculate
the mean and scatter of the resulting 64 autocorrelation estimates.
(We have taken care to eliminate edge effects when calculating
correlations for galaxies within these subvolumes.)  Cyan symbols show
the mean calculated in this way, while the error bars give the $10\%$
to $90\%$ range.  Our models were not tuned to match observational
measures of galaxy clustering, so the overall agreement with
observations demonstrated by Fig.\ref{fig:cf_k} is encouraging.

We can also see that results from the three models converge on large
scales ($\gtrsim 6 h^{-1}{\rm Mpc}$).  On intermediate scales ($6
h^{-1}{\rm Mpc} \gtrsim r \gtrsim 1 h^{-1}{\rm Mpc}$), models B3 and
C3 exhibit weaker clustering than model A1, agreeing better with the
observational estimates.  Interestingly, the 2-point correlation
function of subhaloes (see Fig.~\ref{fig:cf_dm}) shows a similar
offset between the models on these scales. On smaller scales ($r< 1
h^{-1}{\rm Mpc}$) the three models give different results -- B3 agrees
with the observational data to within the scatter found among the 64
Millennium samples, while C3 is marginally high and A1 is
significantly high. Clearly clustering on these scales is quite
sensitive to details of the galaxy formation physics.

It is also interesting that model A1 is significantly high compared to
the mean of the Millennium results on scales between $2$ and $8
h^{-1}{\rm Mpc}$, even though the two simulations adopt the same
galaxy formation physics within the same cosmology.  This suggests
that the particular realisation of a $L= 125 h^{-1}{\rm Mpc}$ box used
in this paper overestimates clustering on these scales.  Correcting
for this would bring models B3 and C3 into excellent agreement with
the observations for $r > 2 h^{-1}{\rm Mpc}$. In summary, all three
models agree well with the data on the large scales that are sensitive
to 2-halo correlations.  Model B3 also agrees well with observation on
smaller scales which are dominated by galaxy pairs within a common
halo. C3 is slightly high on these scales and A1 is too high to be
compatible with the observational data. We note, however, that there 
is a significant difference between B3 and C3 which are implemented 
on the same WMAP3 simulation. This emphasizes that
small-scale galaxy correlations are very sensitive to details of the
adopted galaxy formation physics and are unlikely to be useful for
constraining cosmological parameters.
 
\begin{figure}
\bc
\resizebox{8cm}{!}{\includegraphics{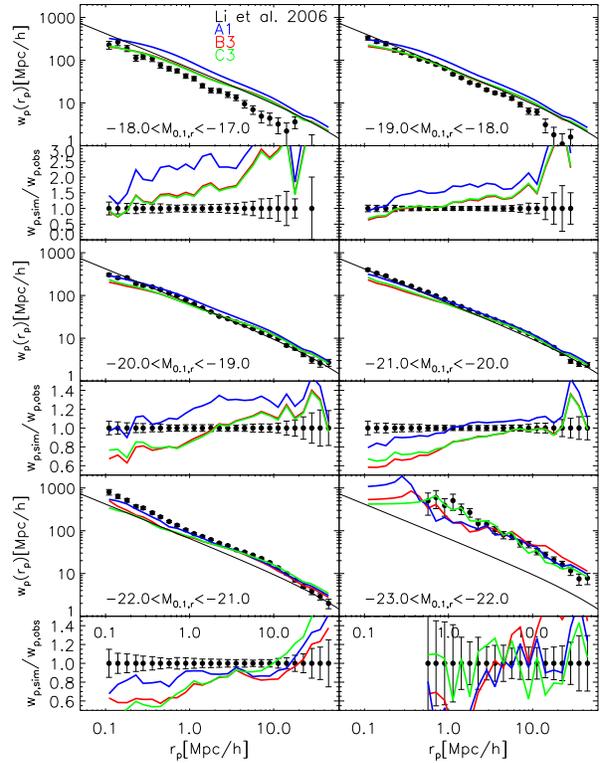}}\\%
\caption{Projected correlation function $w_{\rm p}$ in the SDSS
$r_{\rm 0.1}$ band. Coloured lines show results for our simulated
galaxy catalogues.  Black points are for the SDSS data
\citep{li06}. Each pair of panels corresponds to a different absolute
magnitude bin.  The solid black line in the upper panel of each pair
corresponds to the power-law: $\xi = (r/5\,{\rm
Mpc}\,h^{-1})^{-1.8}$. The lower panel of each pair plots the ratio of
model to observation, with error bars to indicate the uncertainty in
the observational estimate.}
\label{fig:wrp}
\ec
\end{figure}

In Fig.~\ref{fig:wrp} we study how galaxy correlations vary with luminosity,
comparing the projected autocorrelations $w(r_{\rm p})$ from our models (solid
coloured lines) with observational data from the SDSS \citep{li06} (symbols
with error bars).  Results are shown for six magnitude bins from faint (top
left) to bright (bottom right). Below each panel, we also show the ratio
between the model and the observed estimates.  The model projected correlation
function has been determined by integrating the real space correlation
function ($\xi(r)$) along the line-of-sight:
\begin{equation}
w(r_{\rm p})=2~\int_0^{\infty}\xi({\sqrt{r_{\rm p}^2+r_{_{||}}^2}})dr_{||}=2~\int_{r_{\rm p}}^{\infty}\xi(r)\frac{r~dr}{\sqrt{r^2-r^2_{\rm p}}}
\end{equation}
We note that if the integration is truncated at $ r=60 {\rm Mpc}/h $
(half the box size of our simulations) the resulting projected correlation
function is reliable up to $\sim 10{\rm Mpc}/h$.  Because of the
limited volume of our simulations, the two brightest magnitude bins
contain only a few thousand and a few hundred galaxies respectively.
Results for the fainter bins are based on much larger numbers of
galaxies.  The straight black line reproduced in each panel to facilitate
comparison corresponds to the power-law:
\begin{displaymath}
\xi = (r/5\,{\rm Mpc}\,h^{-1})^{-1.8}
\end{displaymath}

The results in this figure show reasonably good agreement between the
models and the observations for $M_r < -20$, but significant
overpredictions, particularly for model A1, at fainter absolute
magnitudes and at large scales. The differences between model and
observation show similar trends with pair separation for all the
models, and are as large as the differences between the models
themselves. On the basis of this comparison none of the models is
obviously better or worse than the others.  In particular, at bright
magnitudes ($M_r<-20$) model A1 is a significantly better fit to the
observations than either B3 or C3. This reproduce the trends which 
\citet{li07} found in their own comparison of SDSS correlation to the 
Millennium Simulation catalogue of \citet{croton06}. This is the 
exact opposite of the conclusion drawn above from Fig.~\ref{fig:cf_k}, 
suggesting that the level of agreement between different observational 
estimates of galaxy correlations is not yet good enough to distinguish 
between the various models we are discussing.

\subsection {Evolution to high redshift}
\label{highredshift}

\begin{figure}
\bc
\hspace{-0.8cm}
\resizebox{8cm}{!}{\includegraphics{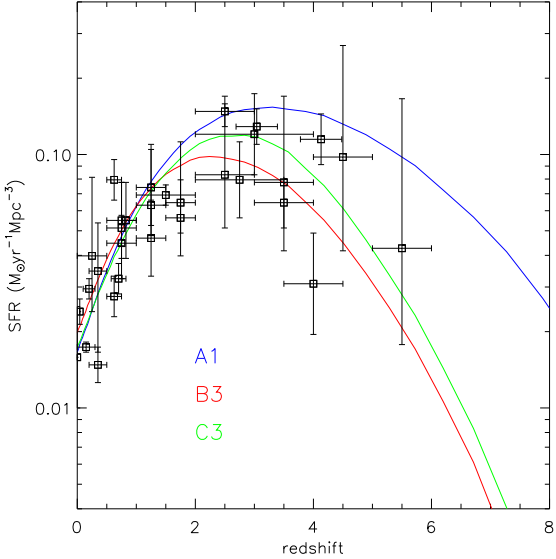}}\\%
\caption{Cosmic star formation rate density as a function of redshift.  Symbols
  with error bars are a compilation of observational data \citep[taken from
  Fig.~12 of][]{springel03}. The solid coloured curves show results from our
  `best' models (see text for details).}
\label{fig:sfr}
\ec
\end{figure}

Figure~\ref{fig:sfr} shows how results from our models compare to the
observed cosmic star formation rate density as a function of redshift.
Symbols with error bars are a compilation of observational data taken
from Fig.~12 of \citet{springel03}.  Models B3 and C3 have quite
similar star formation histories, although C3 lies above B3 by
about $15$ per cent at $z>2$.  This is due to the higher star
formation efficiency used in model C.  Model A1 provides much larger
star formation rates than either B3 or C3 for $z>1$ and lower star
formation rates in the local Universe.  The high redshift difference
reflects the earlier formation of structure in the WMAP1 cosmology,
while the difference at late times is a consequence of the requirement
that all models produce the same total number of stars (as measured by
the K-band luminosity function) in the present universe.

The most dramatic difference in Fig.~\ref{fig:sfr} is that between the
two cosmologies at the highest redshifts.  For $z\sim 6$, the star
formation rates are almost an order of magnitude lower in models B3
and C3 than in model A1. A related result, also visible in 
Fig.~\ref{fig:sfr}, is that the peak of the cosmic star formation 
history is shifted to lower redshift in the new cosmology: 
from $z\sim3$ in model A1 to $z\sim2$ in model B3 or to $z\sim2.5$ 
in model C3.  Fig.~\ref{fig:sfr} suggests that measurements of the 
cosmic star formation rate at high redshift can potentially 
constrain models like those discussed here. Unfortunately, 
observational uncertainties (e.g. due to the use of different star 
formation estimators at different redshifts, and to the need for
substantial dust corrections) are too large to discriminate reliably
between our three models. In addition, these models are far from exhausting
all physically plausible possibilities for the phenomenology of star
formation and feedback, so the true theoretical uncertainty
is undoubtedly larger than suggested by Fig.~\ref{fig:sfr}.

\begin{figure}
\bc
\hspace{-0.8cm}
\resizebox{9cm}{!}{\includegraphics{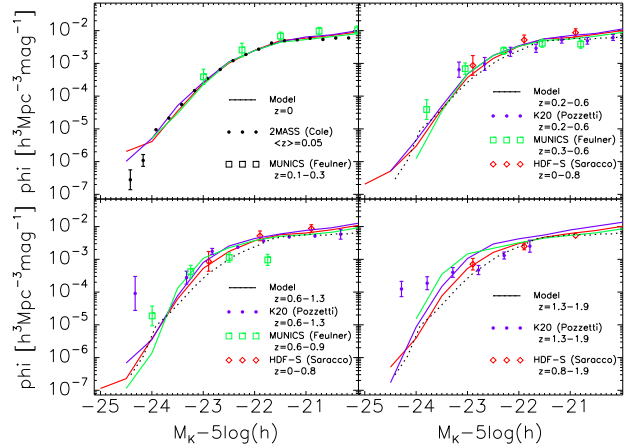}}\\%
\caption{Rest-frame K-band luminosity functions for different redshift intervals.
  Model results are shown as solid lines (blue for A1, red for
  B3, and green for C3).  Symbols with error bars show
  observational estimates from several surveys, as labelled in each panel.
  The low-redshift observational determinations of \citet{cole01} is
  repeated as a dotted black line in the other panels.}
\label{fig:lf_k_evol}
\ec
\end{figure}
\begin{figure*}
\bc
\hspace{-0.8cm}
\resizebox{18cm}{!}{\includegraphics{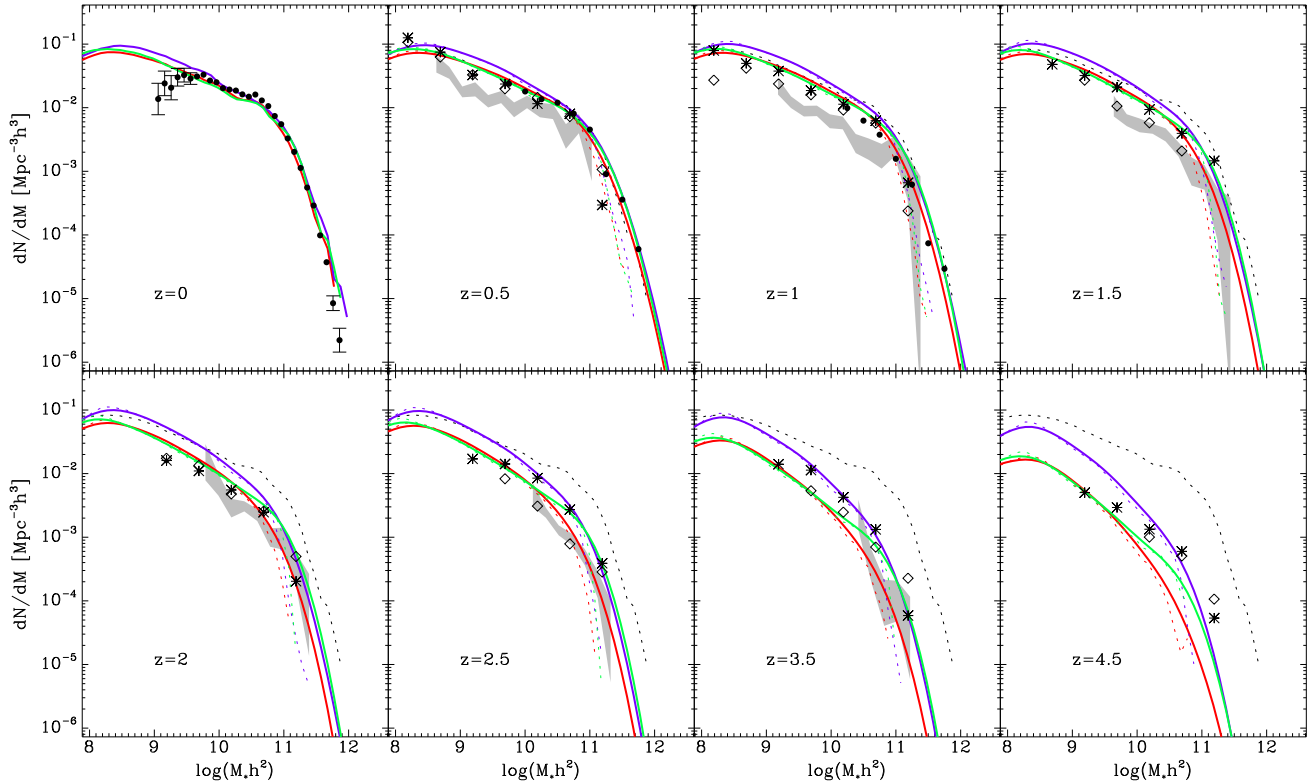}}\\%
\caption{Evolution of the galaxy stellar mass function from $z=0$ to
$z=4.5$.  Model results are shown as solid lines (blue for A1, red for
B3, and green for C3). Local data are from \citet{cole01} and are
repeated as a black dashed curve in the high redshift panels.  High
redshift data are taken from Drory et al. (2005, symbols) and Fontana
et al. (2006, grey shaded areas). Model predictions are shown both
with (solid) and without (dotted) convolution with a normal
distribution of standard deviation 0.25.  At $z =0$ we consider the
mass determinations precise enough to neglect this effect.}
\label{fig:SMF_evol}
\ec
\end{figure*}

\begin{figure*}
\bc
\hspace{-1cm}
\resizebox{18cm}{!}{\includegraphics{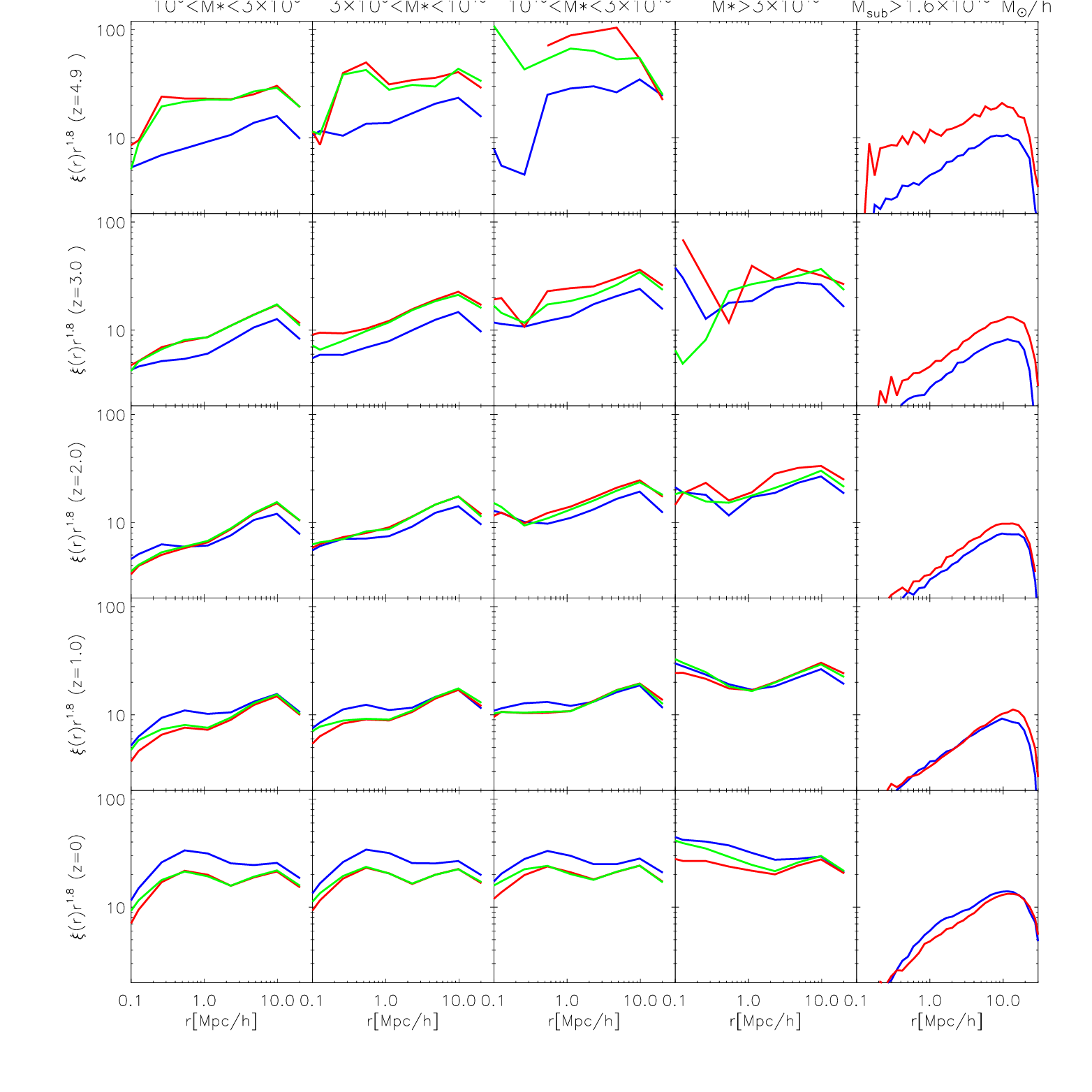}}\\%
\caption{ Autocorrelation functions for galaxies and for resolved
subhaloes at five different redshifts ($z=4.9; 3.0; 2.0; 1.0;
0.$). The galaxy results are given for four disjoint ranges in stellar
mass, as indicated. (Stellar mass $M*$ is given in units of
$M_{\odot}$.) For $M*>3\times10^{10}$ there are too few galaxies at
$z=4.9$ to get meaningful results, so we leave this panel
blank. Different colours in the galaxy panels refer to our different
formation models, blue for A1, red for B3, and green for C3. In the
subhalo column the colours refer to our two simulations, blue for
WMAP1, red for WMAP3.  We have multiplied all correlation functions by
$r^{1.8}$ to make the differences between the models more visible.
Note that $r$ is a comoving coordinate with units of $h^{-1}$Mpc.}
\label{fig:cfhighz}
\ec  
\end{figure*}

In a recent paper \citet{kitzbichler07} compared results from the
model discussed in \citet{delucia07} to a variety of observational
data at high redshift.  They found this model (which is identical to
our model A) to give moderately good agreement with the observed
luminosity and stellar mass functions of galaxies over a large
redshift range. As shown in Fig.~\ref{fig:sfr}, the three models used
in the present study have a significantly different behaviour at high
redshift, so it is interesting to see if the observations can
discriminate between them.

Our figures~\ref{fig:lf_k_evol} and \ref{fig:SMF_evol} correspond to
figures~5 and 7 of \citet{kitzbichler07} and show evolution with
redshift of the rest-frame K-band luminosity function, and of the
stellar mass function respectively.  In Fig.~\ref{fig:lf_k_evol},
symbols with error bars show observational determinations from
\citet{cole01,pozzetti03, feulner03,saracco06}.  The observational
estimate in the local Universe from \citet{cole01} is repeated in the
other panels as a black dashed line.  Model results are shown as solid
coloured lines (blue for A1, red for B3 and green for
C3). Fig.~\ref{fig:lf_k_evol} already showed all three models to agree
nicely with local observations.  At higher redshifts, the agreement is
also reasonably good.  All three seem to underpredict the number of
luminous galaxies in the highest redshift bin, but it should be kept
in mind that the rest-frame K-band luminosities here have been
extrapolated beyond the directly observed region, and so are quite
uncertain \citep[see the discussion in][]{kitzbichler07}. Our three
models start to show significant differences for
$z>1$. Note that model A1 lies between models B3 and C3 in this plot
so that the differences are due mainly to galaxy formation physics
rather than to cosmological parameters.  They are, in any case,
comparable to the uncertainties in the observations, so that
significantly better data are required at these redshifts to put
strong constraints on our models.  Finally, we note that the error
bars of Fig.~\ref{fig:lf_k_evol} underestimate the true uncertainties
as they do not include the effects of cosmic variance. These are
particularly important when small regions of the sky are sampled - as
is the case for the data in the highest redshift bin.

Fig.~\ref{fig:SMF_evol} shows the evolution of the galaxy stellar mass
function for our three ``good'' models (coloured solid and dashed
lines) and compares them with observational determinations from
\citet{cole01,drory05,fontana06}.  As in \citet{kitzbichler07}, model
results are shown both with (solid) and without (dashed) convolution
with a normal distribution of standard deviation 0.25 dex, intended to
represent measurement errors in ${\rm log}\,{\rm M}_*$.  The dashed
black line in each panel repeats the local observational estimate by
\citet{cole01}. At redshifts beyond 1, model A1 predicts substantially
more galaxies in the mass range $9 < \log (M_*/M_\odot) < 10.6$ than
either of the models in the WMAP3 cosmology. This remains true at
higher redshift for model B3, but not for model C3. For $\log
(M_*/M_\odot) > 10.8$ models A1 and C1 predict similar numbers of
massive galaxies at all redshifts. The increased star formation
efficiency in model C as compared to model A clearly compensates for
the lower $\sigma_8$ in the WMAP3 cosmology. This
demonstrates that even at high redshift it may be difficult to use
galaxy data to distinguish between cosmologies unless the physics of
galaxy formation can be independently constrained. Model B3 severely
underpredicts the number of massive objects at $z=4.5$, but even this
disagreement may not be significant once cosmic variance and
observational uncertainties are taken into account.

Finally, we look at the evolution of clustering to high
redshift. Fig.~\ref{fig:cfhighz} presents spatial 2-point correlation functions
at five different redshifts for all resolved subhaloes (for comparison with the
lower panel of Fig.~\ref{fig:cf_dm}) and for galaxies in four different stellar
mass bins in each of our three ``good'' galaxy formation models. Perhaps
surprisingly, although the result of section~\ref{sec:dm}, that the clustering
of resolved subhaloes is very similar in our two cosmologies, is even more
accurately true at $z=1$, it does not hold out to high redshift.  For $z>3$,
resolved subhaloes are actually substantially {\it more} strongly clustered in
the WMAP3 cosmology than in WMAP1, despite the fact that the former has a
significantly {\it lower} mass clustering amplitude. This effect is also
visible in the galaxy autocorrelations. At the present day our models predict
galaxies of all stellar masses to be somewhat more strongly clustered in the
WMAP1 cosmology than in the WMAP3 cosmology.  Beyond $z=2$ the opposite is
true. By $z=5$ the effect is quite strong, more than a factor of two in
correlation amplitude. These effects may seem surprising, but in fact the
corresponding results for dark halos are easily obtained if standard analytic
models are applied to our two cosmologies.  For the convenience of the reader
we provide an appendix repeating Mo \& White's (2002) graphical analysis of
evolution in the abundance and clustering of halos for these two cases.
The large difference in the predicted galaxy clustering
properties in the two cosmological model could potentially help to diagnose
cosmological parameters. We note, however, that our model is not able to
predict reliably high redshift galaxy populations such as Ly-$\alpha$ emitters,
ultraluminous infrared galaxies etc. In addition, observational
measurements at these redshifts are affected by systematics that are not
currently well understood. A rigorous comparison between model predictions and
observational measurements at these redshifts is thus still difficult.

\section{Conclusions and Discussion}
\label{sec:conclusions}

We have carried out cosmological structure formation simulations of a
$\Lambda$CDM Universe for the cosmological parameter sets suggested by
the first- and third-year WMAP results.  The significant reduction in
the best value for the amplitude of matter fluctuations on
$8\,h^{-1}\,{\rm Mpc}$ scale ($\sigma_8$) combines with the decrease
in the estimate of the scalar spectral index for primordial
perturbations ($n$), and with the lowered matter density ($\Omega_m$)
to produce a significant delay in structure formation in the WMAP3
case (see Sec.~\ref{sec:dm}).

By coupling our numerical simulations to semi-analytic models for
galaxy formation, we have investigated the implications of this delay
for the observed properties of galaxies, both at low and at high
redshift.  Specifically, we have compared the galaxy formation model
described in \citet{delucia07} for the WMAP1 cosmology to two galaxy
formation models for the WMAP3 cosmology which use the same physical
framework but different efficiency parameters.  We find that both new
parameter sets can compensate for the delay in structure formation to
produce galaxy populations at $z=0$ which agree with observation just
as well as the old model for the WMAP1 cosmology. The luminosity
functions are almost identical, the correlation functions show at most
small differences, and offsets in the predicted Tully-Fisher
relations are difficult to interpret because disk rotation velocities
cannot be predicted reliably to the level of accuracy required.

Pairwise velocity dispersion measurements are sensitive to cluster
  abundance and therefore differ significantly for the two cosmological models
  of our study. We have shown that the two WMAP3 models underpredict the
  measured PVD by more than $\rm  100 km/s$ on scales $\rm 0.3<r_p<2 Mpc/h$, and are
  lower than the corresponding predictions of the WMAP1 model by $\rm 50
  \sim 150 km/s$ on all scales.

Substantial differences between the various models appear at high
redshift.  The delay in the structure formation translates directly
into a delay in the global star formation history: at $z\sim 6$ the
star formation rates in the models based on WMAP3 are lower than those
based on WMAP1 by almost an order of magnitude.  As discussed
elsewhere, this has important implications for the formation of the
first stars, and for reionization.  Predictions of our three models
for galaxy luminosity and mass functions at high redshift show
substantial differences. Unfortunately, the uncertainties in the
observed luminosities and masses, combined with large cosmic variance
uncertainties, are still too large to place strong constraints on the
efficiencies and scalings of the physical processes we model. Somewhat
counterintuitively, we find that at high redshift galaxies
of given stellar mass are predicted to be substantially more
clustered in the WMAP3 cosmology than for WMAP1.
 
When comparing $z=0$ correlation functions from our three models to
recent observational determinations from the 2dFGRS \citep{hawkins03}
and the SDSS \citep{li06}, we found interesting and apparently
significant differences between the models, particularly for galaxies
fainter than $\sim -20$ in the SDSS r-band.  At these magnitudes, our
WMAP3 galaxy catalogues provide better agreement with the SDSS data
but the opposite is true for galaxies around the knee of the
luminosity function.  It is important to realise that the reduced mass
clustering amplitude implied by the WMAP3 parameters is almost
entirely offset by an increase in halo bias, so that predictions for
galaxy clustering change very little.  At least at separations $r < 20
h^{-1}{\rm Mpc}$, galaxy clustering is much more sensitive to galaxy
formation physics than to cosmological parameters \citep[see also the
discussion in][]{vandenbosch03}.  In fact, for almost all of the
population properties we have examined, the variations induced by
``acceptable'' variations in the galaxy formation parameters are at
least as large as those produced by the variation in cosmological
parameters between WMAP1 and WMAP3. The systematic properties of
galaxies and their small-scale clustering should be used to understand
how galaxies form, not to constrain cosmology.  A similar conclusion
was reached by \citet{kauffmann99a} who used cruder semianalytic
models to show that similar $z=0$ clustering was predicted in two
quite different cosmological models ($\tau$CDM and $\Lambda$CDM). In
this case, however, the predicted evolution to high redshift was
different enough to offer a clear way to distinguish the models
\citep{kauffmann99b}.

In this paper we have shown that varying efficiency parameters within
a given framework for modelling galaxy formation can lead to very
similar predictions for the evolution and clustering of galaxies in
the WMAP1 and WMAP3 cosmologies. The problem is sufficiently
degenerate that a variety of acceptable parameter sets can be found in
either cosmology. It may seem unsatisfactory to `fine-tune' model
parameters to fit the observational data, but it is interesting that
substantially different efficiencies of star formation and feedback
are required in the various cases. More detailed observational data on
how these processes work in individual systems may therefore shed
light on which parameter values are appropriate. In addition,
differing efficiencies translate into significantly different
predictions at high redshift. More detailed and more statistically
precise observations of high-redshift galaxies will be able to
distinguish between the models. Once these aspects of the galaxy
formation process are better understood, it may indeed be possible to
use galaxy surveys to constrain cosmological parameters.

\section*{Acknowledgements}
We thank Volker Springel for providing us with the simulation code
{\small GADGET}2 and with post-processing software.  We thank Cheng Li
for help in understanding the SDSS data and for
providing the PVD fitting code used in his paper. We thank Lan Wang
for providing the HOD galaxy catalogue from her Millennium Simulation
model. We also thank  J.~Blaizot, L.~Gao, G.~Guzzo and Y.~P.~Jing for
helpful discussions, and an anonymous referee for suggesting the
inclusion of the PVD analysis and offering other useful comments.  The
simulations described in this paper were carried out on the Blade
Centre cluster of the Computing Center of the Max-Planck-Society in
Garching. The Millennium Simulation data used in this paper are
publicly available from http://www.mpa-garching.mpg.de/millennium
\label{lastpage}
\bibliographystyle{mn2e}
\bibliography{w3y_v3}

\appendix 
\section {Halo abundance and clustering}
\label{appendix1}

\begin{figure}
\bc
\hspace{-0.8cm}
\resizebox{8cm}{!}{\includegraphics{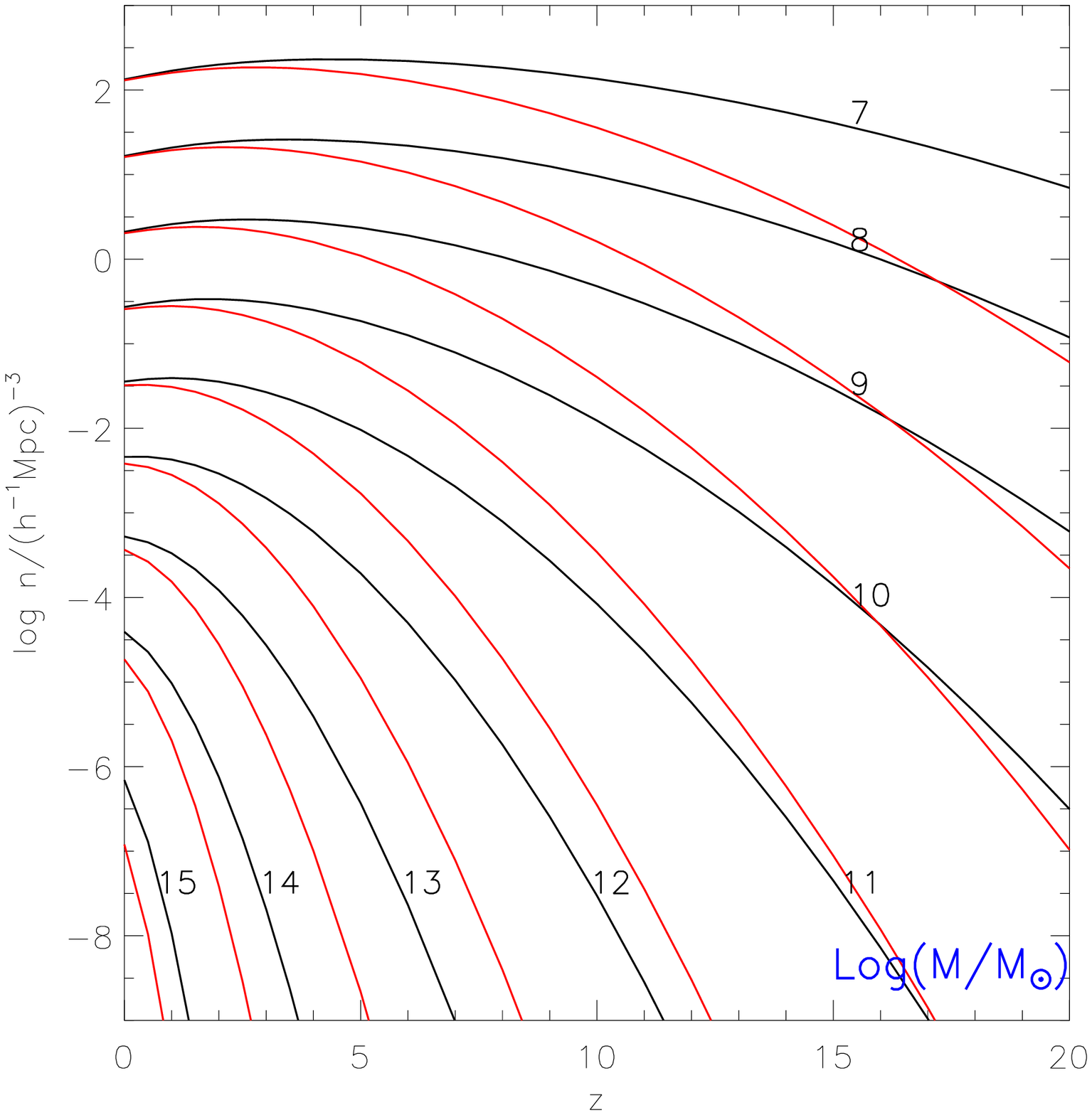}}\\%
\caption{The evolution of the comoving number density of dark matter
haloes with mass exceeding a specific value M in our two cosmological
models: WMAP1 (black curves) and WMAP3 (red curves). The numbers labelling the
black curves indicate the value of $\rm log(M/M_{\odot})$ for the WMAP1 case. 
See the text for more details.}
\label{fig:abundance1}
\ec
\end{figure}

\begin{figure*}
\bc
\hspace{-0.8cm}
\resizebox{16cm}{!}{\includegraphics{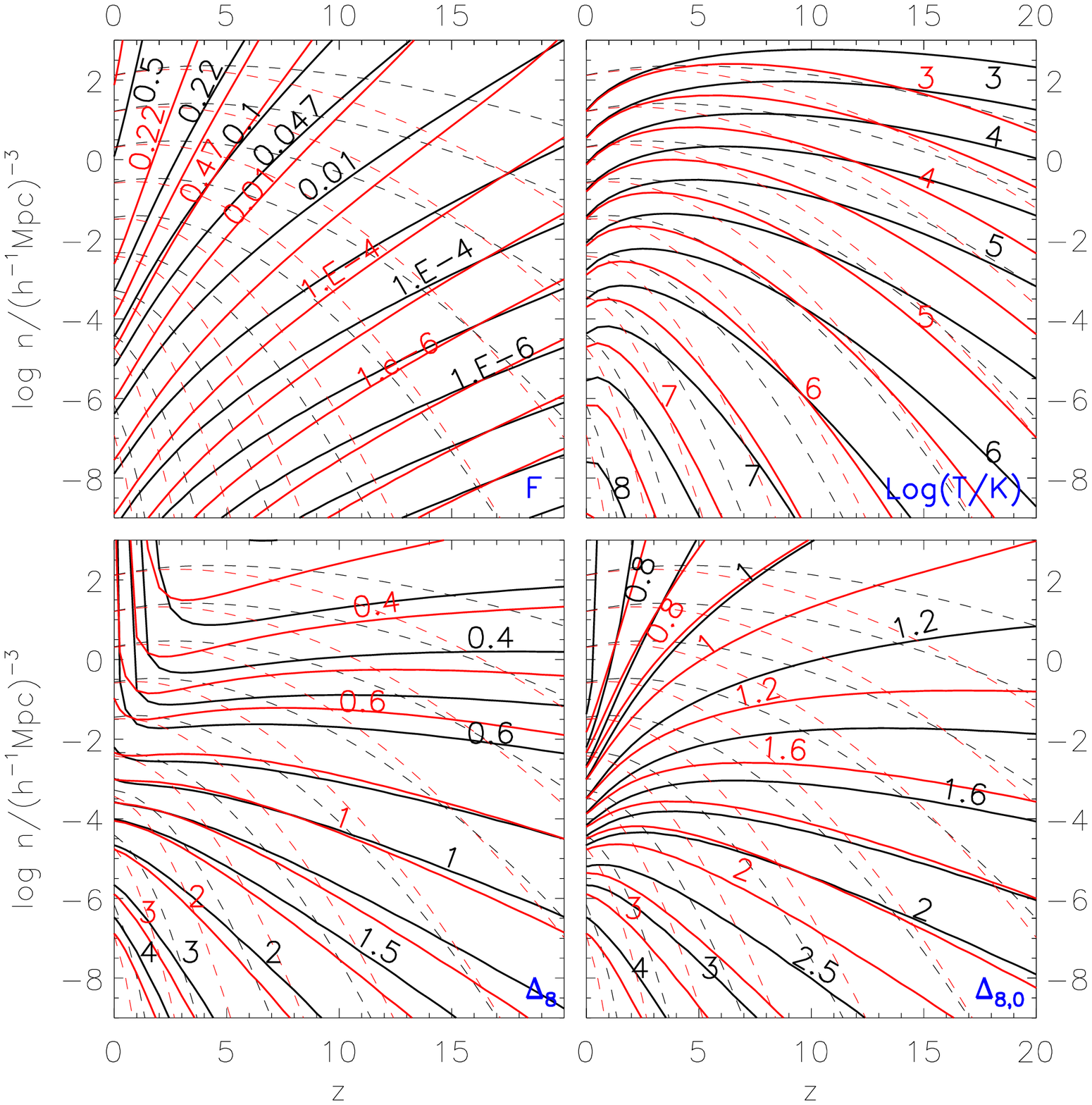}}\\%
\caption{A repeat of figure~2 of \citet{mo02} but showing results for both
  WMAP1 (black) and WMAP3 (red) parameters. The limiting mass of the halo
  population is chosen so as to keep a different quantity constant along a
  solid curve in each panel: cosmic mass fraction $F$ (top left), minimum
  virial temperature $T$ (top right), clustering strength $\Delta_8$ (bottom
  left) and clustering strength of the $z=0$ descendents $\Delta_{8,0}$
  (bottom right). The dashed curves in each panel repeat those of
  Fig.~\ref{fig:abundance1}. The black and red numbers
  label the WMAP1 and WMAP3 curves, respectively. See the text for details. }
\label{fig:abundance2}
\ec
\end{figure*}

In this appendix we use the formulae and the graphical presentation scheme of
\citet{mo02} to illustrate how the evolution of halo abundance and clustering
differs between our WMAP1 and WMAP3 cosmologies.  As in figures 1 and 2 of
\citet{mo02}, we present plots of comoving abundance against redshift for halo
samples defined by lower mass limits $M_{min}(z)$ corresponding to a variety
of halo properties, in particular, for lower limits which correspond at all
redshifts to a given halo mass, a given halo virial temperature, a given
fraction of the total cosmic mass density, a given clusering strength in
comoving units, and a given clustering stgrength at $z=0$ for the halo
descendents. We refer to \citet{mo02} for detailed discussion of these
quantities and for the relevant formulae.

Figure~\ref{fig:abundance1} corresponds to figure~1 of \citet{mo02}. Each
curve gives the comoving abundance as a function of redshift of halos more
massive than the value indicated by the label. Curves for our WMAP1 parameters
are shown in black while curves for our WMAP3 parameters are shown in red.
Labels give the decimal logarithm of halo mass in units of solar
masses and are placed next to the corresponding WMAP1 curve. Curves are
plotted for the same set of mass limits in the two cosmologies, and give
almost identical abundances at $z=0$ for low masses.  This can be used to
identify the WMAP3 curves at high redshift where they deviate very
substantially from the WMAP1 curves. \citet{springel05a} showed these
theoretical predictions to be in excellent agreement with Millennium Simulation
results for $z\leq 10$. At all masses the difference in halo abundance 
between the two cosmologies increases with the increasing redshift. For
$M> 10^{15}M_{\odot}$ the abundance difference is already almost
an order of magnitude at $z=0$, and the same is true for $M>
10^{12}M_{\odot}$ at $z=5$ and $M> 10^{8}M_{\odot}$ at $z=10$.

In Fig.~\ref{fig:abundance2}, we give abundance-redshift relations for halo
samples defined above lower mass limits $M_{min}(z)$ which correspond to other
conditions, as in figure~2 of \citet{mo02}. In each of these plots the two
curves of Fig.~\ref{fig:abundance1} are repeated as dashed curves.  These can
be used as a reference to obtain the halo mass corresponding to each point in
the abundance-redshift plane.

In the upper left panel of Fig.~\ref{fig:abundance2} the solid curves link
halo populations containing given fractions $F$ of the total cosmic mass
density at each redshift. The labels give $F$ values for the 
curves they are placed next to.  Where no label is given the $F$ value can be
inferred from the surrounding curves. Red curves show WMAP3 results for the
same $F$ values and lie above the corresponding WMAP1 curves at all redshifts.
At $z=0$ the curves are close enough that it is easy to infer the $F$ value
for each red curve by comparing it with the corresponding black curve. For
example, for WMAP1 at $z=0$, one percent of all cosmic mass is in dark halos
above a lower mass limit corresponding to abundance $n = 5\times
10^{-7}h^3{\rm Mpc}^{-3}$, thus $M>10^{15}{\rm M}_\odot$ (from
Fig.~\ref{fig:abundance1}). For WMAP3, the $z=0$ abundance at $F=0.01$ is
about a factor of 2 higher, and the corresponding mass is about 2.5 times
smaller.  At $z=5$ the one percent mass point corresponds to $n =
10^{-3}h^3{\rm Mpc}^{-3}$ and $M>4\times 10^{11}{\rm M}_\odot$ for
WMAP1, but to $n = 10^{-2}h^3{\rm Mpc}^{-3}$ and $M>4\times 10^{10}{\rm
M}_\odot$ for WMAP3.

In the upper right panel of Fig.~\ref{fig:abundance2} the solid curves link
halo populations at each redshift with virial temperatures $T$ in excess of a
given value. Labels give the decimal logarithm of the limiting temperature in
Kelvin and are placed next to the curve they refer to. At $z=0$, there
is a close correspondance between WMAP1 and WMAP3 at high abundance. At low abundance
(high mass) the WMAP3 curves lie below their WMAP1 counterparts.  At $z=10$,
the current best estimate of the reionization redshift, only halos with
$M>8\times 10^{7}{\rm M}_\odot$ have virial temperatures sufficient to
ionize hydrogen ($T> 10^4K$) and are thus able to cool their baryonic
component effectively. For WMAP1 parameters, the comoving abundance of such
halos is $n = 10h^3{\rm Mpc}^{-3}$ and they contain a fraction $F\sim 0.04$ of
all cosmic matter (from the upper left plot of Fig.~\ref{fig:abundance2}). For
WMAP3 parameters, the predicted abundance of such halos drops by about a factor
of 6, and the fraction of cosmic matter contained in them drops by about an
order of magnitude.  This is the reason why reionization is much more
difficult to explain (at $z=10$) for the revised WMAP parameters. Differences
of this kind also explain why we get higher global star formation rates at
$z\sim 5$ in our model A1 than in our models B3 and C3 (Fig.\ref{fig:sfr}) as
well as a correspondingly higher stellar mass function in A1 at high redshift
(Fig.~\ref{fig:SMF_evol}).

In the lower left panel of Fig.~\ref{fig:abundance2} the solid curves link
halo populations at each redshift which have a given strength of clustering in
comoving coordinates as characterized by $\Delta_8 (M,z)$, the rms fluctuation
in overdensity of haloes more massive than $M$ at redshift $z$ after smoothing
with a spherical top-hat filter of comoving radius $8h^{-1} {\rm Mpc}$. As in
the other panels, the labels give $\Delta_8$ values for the curves they
are placed next to. At redshifts 5 to 10, the WMAP3 curves lie below the
corresponding WMAP1 curves at low abundances but above them at high
abundances. Comparing the dashed and solid curves in this panel, one sees that
at $z=0$ the clustering strength of halos is very similar in the two
cosmologies for all lower limits to halo mass. This is the result visible in
the lower panel of Fig.~\ref{fig:cf_dm}. At $z=5$, however, these same curves
show the clustering strength of halos to be substantially stronger in WMAP3
for all limiting halo masses than in WMAP1 (typically by 25 to 40\% in
$\Delta_8$). This is the result seen for low-mass (sub)halos in the right
column of Fig.~\ref{fig:cfhighz}.

Finally, in the lower right panel of Fig.~\ref{fig:abundance2} solid curves
link halo populations at each redshift for which the $z=0$ descendents have a
given clustering strength, as indicated by $\Delta_{8,0}$, their present-day
value of$\Delta_8$. As in the other panels, the labels give values of
$\Delta_{8,0}$ for the curves they are placed next to. At $z=0$, the two 
cosmologies predict nearly identical clustering strengths at intermediate 
abundances. \citet{mo02} give a number of examples of how this plot may be 
used. Here we note that the progenitor population which ends up with a 
given $z=0$ clustering strength is substantially more strongly clustered at 
high redshift in the WMAP3 cosmology than in the WMAP1 cosmology.

\bsp
\end{document}